\documentclass[showpacs,preprintnumbers,amsmath,amssymb]{article}
\usepackage[textwidth=17cm,textheight=24cm]{geometry}
\usepackage{multicol}
\usepackage{graphicx}% Include figure files
\usepackage{dcolumn}% Align table columns on decimal point
\usepackage{bm}% bold math
\usepackage{wrapfig}
\usepackage{graphicx}
\usepackage{xcolor}
\usepackage[latin1]{inputenc}
\usepackage{graphicx}
\usepackage{graphics}
\usepackage{amsmath}
\usepackage{amssymb}
\usepackage{amsfonts}
\usepackage{mathrsfs}
\usepackage{latexsym}

\renewcommand\d\delta
\newcommand\D\Delta
\newcommand\beq{\begin{equation}}
\newcommand\eeq{\end{equation}}

\usepackage{amsfonts}
\usepackage{relsize}
\usepackage{graphicx}% Include figure files
\usepackage{dcolumn}% Align table columns on decimal point
\usepackage{bm}% bold math
\usepackage{wrapfig}
\usepackage{graphicx}
\usepackage{url,lineno}
\usepackage{amsfonts}
\usepackage{color}
\usepackage[english]{babel}

\begin{document}
\date{}
\title{\bf A predictive model for protein materials: \\ from macromolecules to macroscopic fibers}

\maketitle \vspace{-1.8 cm}
\begin {centering} {\bf G. Puglisi$^1$, D. De Tommasi$^1$, M.F. Pantano$^2$, N. Pugno$^3$, G. Saccomandi$^4$.}
\vspace{0.3 cm}

\noindent{$^1$ Dipartimento di Scienze dell'Ingegneria Civile e dell'Architettura,
Politecnico di Bari, Italy}
\vspace{0.3 cm}

\noindent{$^2$ Laboratory of Bio-Inspired and Graphene Nanomechanics, Department of Civil, Environmental and Mechanical Engineering, 
University of Trento, Via Mesiano 77, 38123 Trento, Italy}
\vspace{0.3 cm}

\noindent{$^3$ Laboratory of Bio-Inspired and Graphene Nanomechanics, Department of Civil, Environmental and Mechanical Engineering, 
University of Trento, Via Mesiano 77, 38123 Trento, Italy;
\\ Agenzia Spaziale Italiana, Via del Politecnico, 00133 Roma (Italy);
\\School of Eng. and Mat. Science, Queen Mary University of London, Mile End Road, London E1 4NS, U.K.}
\vspace{0.3 cm}

\noindent{$^4$ Dipartimento di Ingegneria, Universit\`a degli Studi di Perugia. \\ School of Mathematics, Statistics and Applied Mathematics,
NUI Galway, University Road, Galway, Ireland}

\date{}

\vspace{0.5 cm}

\begin{abstract}
 \noindent We propose a model for the mechanical behavior of protein materials. Based on a limited number of experimental macromolecular parameters (persistence and contour lengths, rate of unfolding dissipation) we obtain the macroscopic behavior of  keratin fibers (human, cow, and rabbit hair), taking into account the damage and residual stretches effects which are fundamental in many functions of life. We support our theoretical results by showing that our model is robust and able to reproduce with high quantitive accuracy the cyclic experimental behavior of different keratinous protein materials we tested.  We also show the capability  of describing, even if with lower precision, the dissipation and permanent strain effects in spider silks.
\end{abstract}

\vspace{0.5 cm}

\end{centering}

\begin{multicols}{2}

\section{Introduction} 
 Experimental analyses (AFM, optical and magnetic tweezers, nanoindentation) clearly show that  the outstanding elasticity, toughness, strength, and self-healing properties \cite{Bub, ESL} of protein materials originate from their secondary structure,  characterized by the presence of folded (crystal-like) domains, typically in the form of $\alpha$-helices or $\beta$-sheets, which can undergo unfolding as a consequence of an applied displacement.  The efficacy of the unfolding mechanism is based on the presence of a large number of non-covalent forces \cite{Fan}, typically hydrogen-bonds, which act as sacrificial joints. Being much weaker than the covalent peptide bonds, these interactions can easily break,  causing unfolding phenomena  ($\alpha \rightarrow \beta$ transition or $\beta$ domains unfolding). Such unfolding  prevents the backbone fracture conciliating the typically conflicting requests for high stiffness and toughness  \cite{RIT}.
 Indeed, the cooperative strength of non-covalent bonds provides proteins with stability and stiffness before unfolding  begins, whereas the increase of the end-to-end length during unfolding allows to accommodate large stretches (20$\%$-40$\%$), resulting in high energy dissipation and toughness.

 Here we focus on two different classes of structural protein materials \cite{BY} known for their outstanding performance: {\it keratinous materials}, which can be found in wool, hair, cells intermediate filaments, epithelial cells and hooves, and {\it silk}, a valuable material produced by spiders and silkworms. In the first case the secondary structure is in the form of a $\alpha$-helices undergoing a transition to an almost straight polypeptide chain in the form of $\beta$-sheets \cite{Hea, SSH}; whereas in the second case the protein secondary structure is in the form of  (alanine rich) $\beta$-sheet folded domains which undergo unfolding under stretching \cite{OS}. 
 Both silk and keratinous materials have attracted significant attention for the design of new bioinspired materials \cite{ESL, SSH, DNL,  GOS, KY, LK, ZLY, WSK}, such as nano-polymers \cite{LK} and block copolymers \cite{ZLY} in the case of spider silk and hydrogels  \cite{WSK} in the case of  keratinous materials. 
  Protein unfolding occurring in the above listed proteins, but also in others, like cytoplasmic IF proteins \cite{Bu}, needs a deep comprehension as it is linked to many biological phenomena 
 (i.e. mechano-transduction and cell motility \cite{VS}) and human deceases, such as Alzheimer, type II Diabetes, prion diseases, Parkinson  \cite{Se}. 
 
 In this context, as in the case of multiscale dislocation and defect theories in metal plasticity \cite{ZR}, a significant advance of knowledge is strictly related to the availability of models relating the material properties at the macromolecular scale to the macroscopic response of wires and tissues \cite{OS,  CX,  AP,  KB, SSH, ESL}. 
However, modeling complex phenomena, such as  hard domains stability under {\it mechanical loading},  different behavior under thermal and mechanical loading \cite{Paci}, asymmetric unfolding/refolding behavior under cyclic loading, still represents a fundamental open task.  Molecular Dynamics approaches are limited by  the involved time scales   \cite{Makk, Rico}, whereas the possibility of formally connecting  the observed macroscopic rate-independent
dissipation with the micro rate-dependent dissipation 
\cite{Ons} can be provided by considering the evolution in the multi-valley energy landscapes with three different time scales of loading, thermal vibration, and in-well relaxation \cite{KT, PT, Givli}.

 Here, we follow a {\it phenomenological} approach and, as in \cite{KB} and \cite{DMPS}, we consider for the single molecule under assigned elongation a phenomenogical Griffith type approach assuming that the macromolecule unfolds when the energy gain (evaluated as difference of the two force-elongation curves between which the macromolecule `jumps') equals a material parameter, representing the energy `dissipated' in the mechanical unfolding of the crystal. This approach is supported by the experimental observation that the  macroscopic toughness
($\sim 10^2$MJ/m$^3$) of the considered materials corresponds to the energy of the H-bonds of the   unfolded fraction during elongation \cite{Pap}.

 The  contour length variations associated to the unfolding effect can  be similarly evaluated by the macromolecular stretching experiment \cite{RGO}. Then, based on a  (slightly amended) affinity hypothesis \cite{Rubinstein} and the James and Guth three-chains model \cite{WG}, following \cite{DPS} we deduce the macroscopic behavior with damage and residual stretch of protein materials and  its {\it analytical} relation with  few key macromolecular material parameters that can be deduced by AFM experiments. 
 
 To show the ability of the model of reproducing the experimental behavior we cyclically tested three different keratin materials and spider dragline silk.

\section{Micro-macro model}
\begin{figure}[!h]\vspace{-0.30 cm }  \hspace{3 cm}\includegraphics[scale=0.1355]{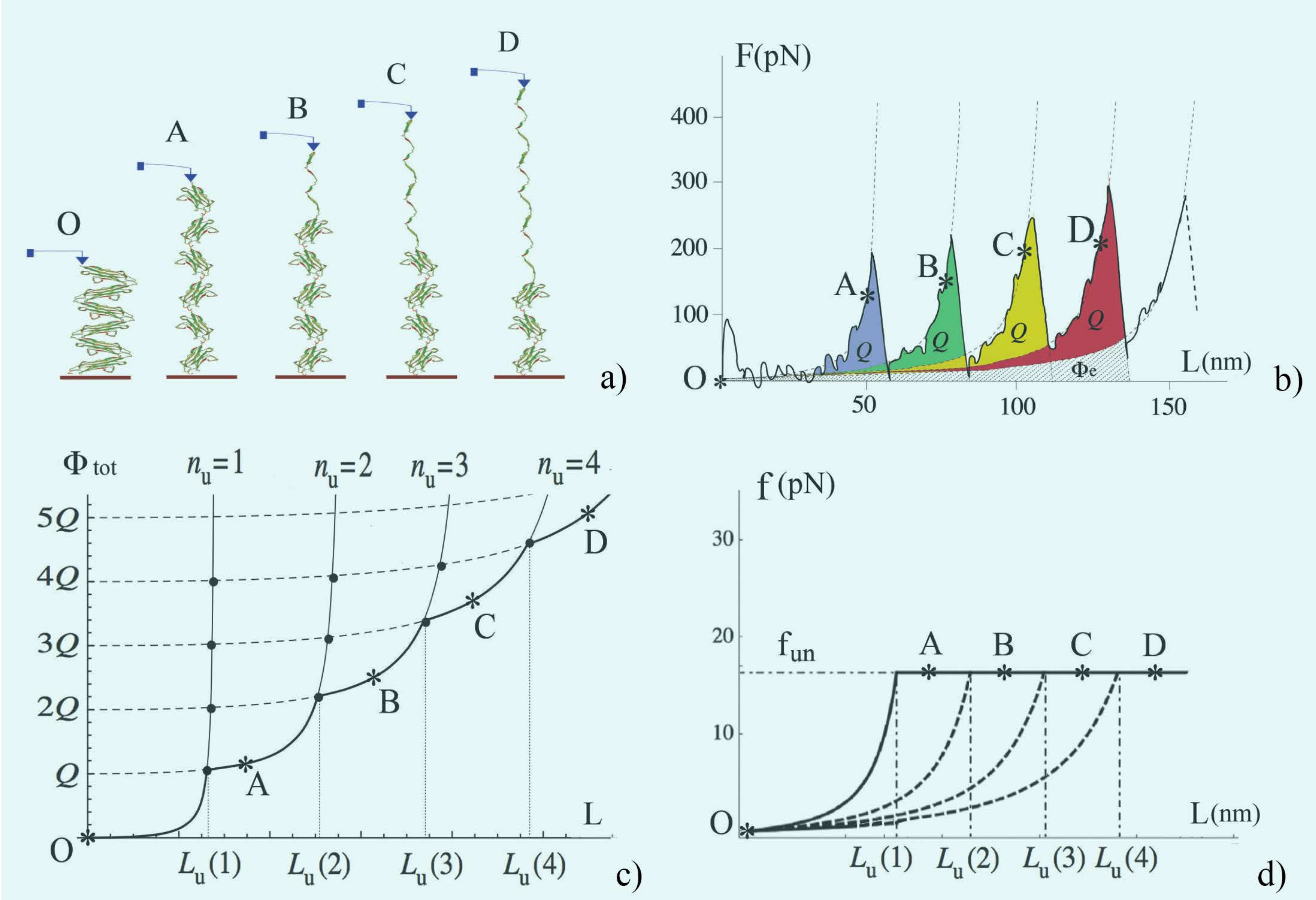}
 \vspace{0.20 cm }  \caption{ \label{single}
a) Scheme of  second structure  unfolding during an AFM molecule stretching and b) corresponding force-elongation curve (reported in \cite{RGO}). The sawtooth shape is due to unfolding of an increasing number of domains with (c) reporting the related dissipated energy (OABCD path). In d) `continuum limit' approximation with constant unfolding plateau}\vspace{-0.3 cm}
\end{figure}

\noindent{\bf Macromolecule energy}. Following a Griffith type total energy minimization approach \cite{DMPS}, as in \cite{Hea} and \cite{Termonia}, we  model a protein macromolecule as a lattice of $n$ two-states (folded/unfolded) domains (Fig.\ref{single}a). We define the state of the $i$-th domain through the variable $\chi_i$ with $\chi_i=0$ $(=1)$ in the folded (unfolded) state. For simplicity,  since we are interested in the unfolding regime, we neglect both the elastic energy and the length of the folded domains. Moreover, we neglect non-local interactions ({\it weak interaction hypothesis}). 
The total internal energy can then be expressed as $\phi_e=\sum_{i=1}^n \chi_i \varphi_e(\lambda_{rel}^i)$, where $\varphi_e$  depends on  the {\it relative deformation} $\lambda^i_{rel}=l_i/l_c$  of the $i$-th unfolded domain, with end-to-end length $l_i$, contour length (assumed constant)  $l_c$ and persistent length $L_p$. 
We also assume the limit extensibility condition: $\lim_{\lambda_{rel}\rightarrow 1}\varphi_e(\lambda_{rel}) =+\infty$.

\end{multicols}

\begin{multicols}{2}
Equilibrium  requires a constant force $l_c \frac{ d \varphi_e(l_i/l_c)}{d l_i}=F $ and,
under the hypothesis of convex $\varphi_e$,
a constant 
$\lambda_{rel}^i=\lambda_{rel}=l_i/l_c=L/L_c$. Here, $L=\sum_i \chi_i l_i$ is the {\it total end-to-end length}  and $L_c=\sum_i \chi_i l_c$
is the {\it total contour length of the unfolded fraction}. As a result we have  $\phi_e=L_c\varphi_e (\lambda_{\mbox{\tiny {\it rel}}})=
 L_c \varphi_e(L/L_c).$ 
Finally, by adopting
the elastic energy density proposed in \cite{DMPS} we obtain $\phi_{e}=\phi_e(L,L_c)=\kappa \frac{L^2}{L_c-L}$, where $\kappa=k_BT/(4L_p)$.

Let us now consider the configurational energy of the different folded/unfolded states.
 Following \cite{DMPS} consider an Ising type transition energy and a di-block approximation (as supported by MD simulations  \cite{HS},
see Supplementary Material, SM, for details), we end up with a  total energy  $\phi_{ tot}=\kappa \frac{L^2}{L_c-L} + \frac{L_c}{l_c} Q$  depending on the single  (history dependent) internal variable $L_c$,  where $Q$ is the unfolding energy of the single hard domain.

In Fig.\ref{single}b we show the scheme of our minimization procedure, requiring minimization of  $\phi_{tot}$ in order to get the transition unfolding length   $L_u$  (see Fig. \ref{single}c and SM for details) corresponding to an increasing number $n_u$ of unfolded domains.  The system stays in a given configuration (fixed folded fraction) until the elastic energy gain equals the unfolding energy $Q$. This energy is then assumed as `dissipated' by heat. The resulting force-end to end length behavior,  shown in Fig.\ref{single}c, reproduces  the experimental behavior in \cite{RGO}  of the single macromolecule. 
\vspace{0.1 cm}

\noindent {\bf Continuum macromolecule approximation}.  Based on the observation of a large number of folded domains (see SM), after introducing the {\it dissipation density}  $q=Q/l_c$,  we may consider a {\it continuum limit approximation} which allows the chain total energy to be rewritten as \begin{equation}\label{Toten}
 \phi_{tot}=\phi_{tot}(L,L_c)=\kappa \frac{L^2}{L_c-L} + q L_c, \end{equation}
 where now $L_c$ can be considered a continuum variable.
The equilibrium force in the continuum case is \begin{equation} \label{df}f=\frac{\partial \phi_{tot}(L,L_c)}{\partial L}= \kappa\frac{2 L \, L_c -L^2}{(L_c-L)^2}.\end{equation}Under our  Griffith fracture hypothesis of global energy minimization ($\partial \phi_{tot}/\partial L_c=0$)  we obtain 
\begin{equation}L_{un}\equiv L_{max}=\frac{L_c}{\sqrt{\kappa/q}+1},  \hspace{0.2 cm} f_{un}=\left(2
   \sqrt{\kappa/q}+1\right)q.
   \label{unf}\end{equation}
As a result {\it the protein unfolding behavior is characterized by a constant stress unfolding plateau, depending only on the persistence length $L_p$ and rate of dissipation $q$} as represented in  Fig.\ref{single}d (path OABCD). 
\vspace{0.1 cm}

\noindent \underline{Remark}.  Dashed lines in Fig. \ref{single}d) are unloading and reloading paths. The introduction of partial refolding upon unloading would let us to describe also internal hysteric effects. Similarly, it would be possible to introduce rate of loading dependent effects \cite{DPSb} upon considering a different macromolecular behavior \cite{Chu}.
\vspace{0.15 cm}

\noindent{\bf Affinity hypothesis}. 
As classical in polymer elasticity \cite{Rubinstein}, the main hypothesis for the deduction of the macroscopic behavior  is the affinity hypothesis which assumes that the macroscopic stretches coincide with the macromolecular ones. 
We here give an interpretation of the permanent deformations observed in protein materials as a macroscopic counterpart of the variation of the macromolecule natural configuration associated with the unfolding of crystal domains and the availability of new monomers ({\it hidden length}). 
We then extend the affinity hypothesis by identifying both permanent and elastic components of the macromolecular stretch with the corresponding macroscopic counterparts.  In such a way we are also able to describe  the important permanent stretch effect.

To analytically describe this phenomenon we begin by observing that for a given contour length $L_c$ the {\it natural length}  (zero force)  $L_n$ of the entropic chain  can be expressed, according to a known result of Statistical Mechanics (see {\it e.g.}\!\!  \cite{Rubinstein}), as 
$$L_n=\sqrt{\bar n} \, b=\sqrt{\frac{L_c}{b}} \, b=\sqrt{L_c \, b} ,$$
where $b$ is the {\it length of the Kuhn segments} and $\bar n$  is the number of Kuhn segments of the unfolded chain fraction. 
Thus, if we denote by $L_c^o$ the initial value of the contour length, the initial natural end-to-end length is
$L_o=\sqrt{L_c^o \, b}.$ Then
we may define the following stretch measures  (see SM): 
\begin{equation}\left \{\begin{array}{ll}\lambda=\frac{L}{L_o},      & \mbox {\underline{total stretch}},\vspace{0.05 cm} \\
\lambda_e=\frac{L}{L_n},& \mbox{\underline{elastic stretch}},\vspace{0.05 cm}\\  \lambda_n=\frac{L_n}{L_o} &  \mbox{\underline{permanent stretch}},\vspace{0.05 cm}\\
 \lambda_c=\frac{L_c}{L_o} &  \mbox{\underline{limit extensibility stretch}}.\label{lam}\end{array}\right . \end{equation}

\end{multicols}

\begin{figure}[!h]\vspace{0.25  cm}
\begin{center}  \includegraphics[scale=0.175]{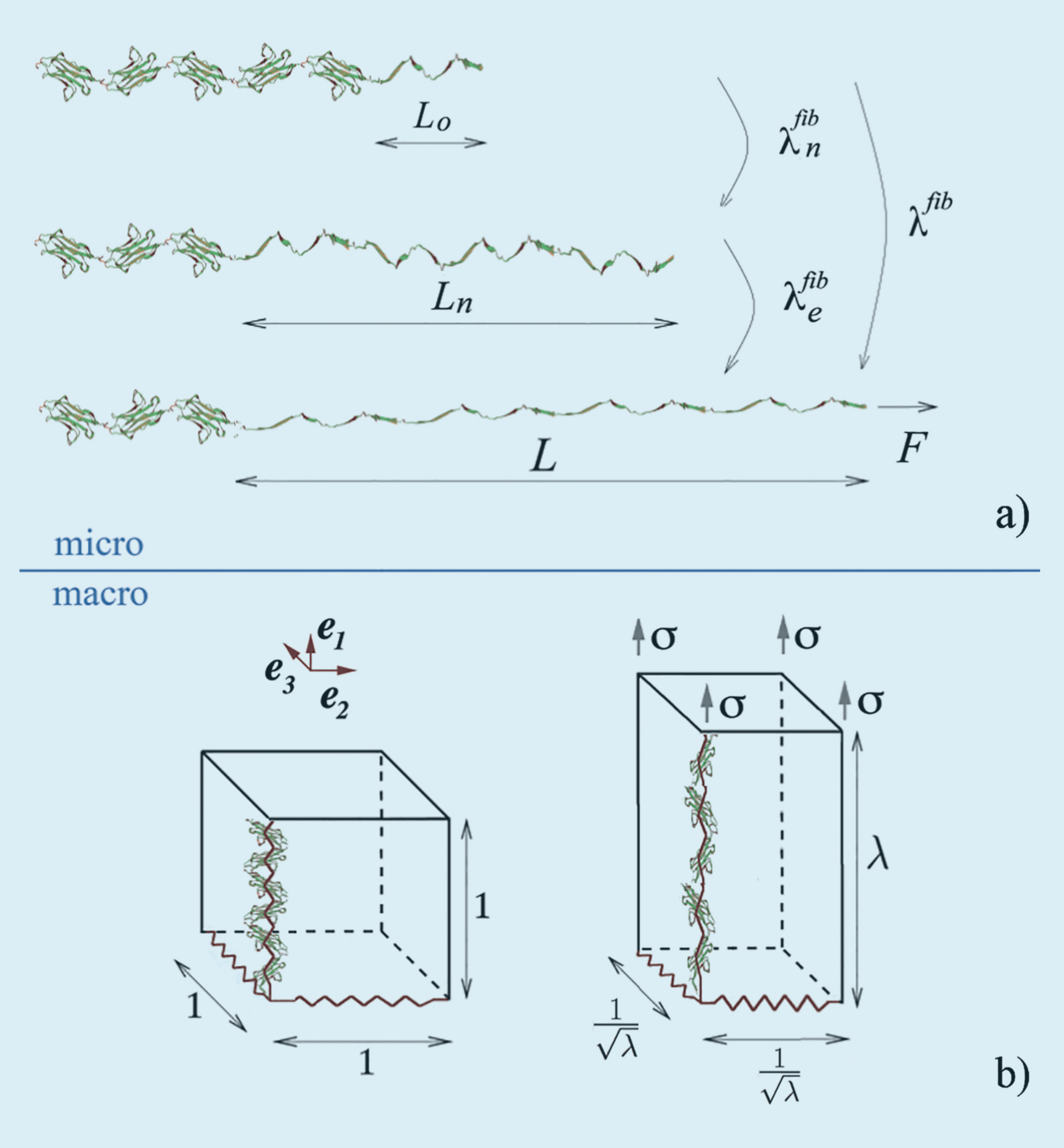}   \caption{ \label{micromacro}
Cartoon of the micro-macro scale transition. a) decomposition of the fibril stretch in elastic and permanent stretch; b) sketch of the extension of a protein assembly in a three-chains scheme: red springs  represent the network Gaussian chains.}\vspace{-0.5 cm}
\end{center}
\end{figure}

\begin{multicols}{2}

\noindent{\bf Macrosopic material behavior}. Structural proteins  are composite materials of hard segments immersed in an amorphous soft fraction of unfolded macromolecules. To describe the complex inter-chains interactions and self avoiding effects, following the  {\it additivity hypothesis} in polymer elasticity \cite{FE}, we determine the energy of the {\it real}  network as the sum of the energy of ideally isolated fibrils (folded molecules in Fig.\ref{micromacro}b)) whose behavior is schematized in Fig.\ref{micromacro}a)) plus a network energy term measuring the chains interactions: red springs in Fig.\ref{micromacro}b). Moreover, according with the classical {\it James and Guth three-chains model}, we assume that all chains are aligned along the three principal macroscopic stretch directions.
 In particular, we suppose that the unfolding protein macromolecules are oriented along the fiber direction, whereas the 
amorphous soft fraction is supposed to be equally distributed along the three principal directions.  For the unfolding chains we assume the continuous limit energy \eqref{Toten}, force \eqref{df}, unfolding forces \eqref{unf}, and stretches \eqref{lam} deduced above.  The network effect is instead modeled by simple Gaussian chains. 

Thus, let $(\lambda_1, \lambda_2, \lambda_3)$ be the macroscopic principal stretches, where $\lambda_1$ is the stretch in the fiber direction. According to previous hypotheses, the macroscopic energy density can be calculated as $\Phi(\lambda_i)=N_{fib}\phi_{tot}(\lambda_1)+\sum_{i=1}^3\frac{N_{net}}{3}\phi_{net}(\lambda_i). $
Here $N_{fib}$ is the number of macromolecules with unfolding domain, per unit  area of fiber section in the reference configuration,     and $N_{net}$ is the number of chains, per unit volume, reproducing the real chains network effect. The network energy
is modeled as Gaussian  $\phi_{net}(\lambda_i)=
\frac{k_B T}{2}
(\lambda_i^2-1)$ leading (\cite{Rubinstein}) to a neo-Hookean network energy $\Phi_{net}=\frac{\mu}{2}(I-3),$
where $I=\sum_{i=1}^3 \lambda_i^2$ is the first invariant of the left Cauchy-Green deformation tensor and $\mu= k_BT N_{net}/3\,$ represents the shear elastic modulus. On the other hand the energy of the unfolding chains gives a history dependent term depending only on $\lambda_1$ stretch. In such a way we describe a fundamental effect with the damage localized along the (maximum elongation) fibre direction. It is worth noticing that this {\it damage anisotropy} is crucial for an effective derivation of the macroscopic permanent stretch.

Suppose now that  the fiber (assumed incompressible) undergoes a {\it simple extension} $\lambda_1=\lambda$ and $\lambda_2=\lambda_3=1/\sqrt{\lambda}$. To take into account the {\it confinement effect}  we introduce a constant pressure $p$ 
 perpendicular to the fiber skin. Thus, if $\sigma=\sigma_1$ is the (Piola/engineering) principal stress in the fiber direction, by using (\ref{df}) and (\ref{lam}) and by imposing the boundary conditions $\sigma_2=\sigma_3=p$, we get the stress-stretch relation \begin{equation}\begin{array}{l}\sigma\!=\!  N^{fib} \kappa \frac{  \frac{2  \lambda}{\bar \lambda_c(\lambda_{max})}-\left ( \frac{\lambda}{\bar \lambda_c(\lambda_{max})}\right )^2}{\left (1- \frac{ \lambda}{\bar \lambda_c(\lambda_{max})} \right )^2}\!+\!\mu^{net}\!\left (\lambda\!-\!\frac{1}{\lambda^2}\right)+\frac{p}{\lambda  \sqrt{\lambda}},\vspace{0.2 cm}\end{array}\label{eeee}
 \end{equation}
where 
$  \lambda_{max}=\lambda \mbox{ if } \dot \lambda_{max} > 0, \mbox{ \it{primary loading} }$
and $\lambda_{max}=\mbox{const}  \mbox{ if } \dot \lambda_{max} = 0  \mbox{ \it{unloading/reloading}. }$
Here $\lambda_{max}$ is the previously maximum attained strain and, by using (\ref{unf}), 
 \begin{equation}\lambda_{c}=\bar \lambda_c(\lambda_{max})=\left (\sqrt{\frac{\kappa}{\bar q}}+1\right ) \lambda_{max}\end{equation}
is the (limit) value of the stretch corresponding to the present value of the contour length.

\section{\bf Experimental validation} 

To prove the efficacy of our model we performed cyclic tensile tests on two different types of material:  keratin hair materials and dragline silk.

\vspace{0.25 cm}

\vspace{0.15 cm}

\begin{figure}[h!] \vspace{-0.4 cm}\hspace{2.5 cm}$$\begin{array}{ccc}
\includegraphics[scale=0.56]{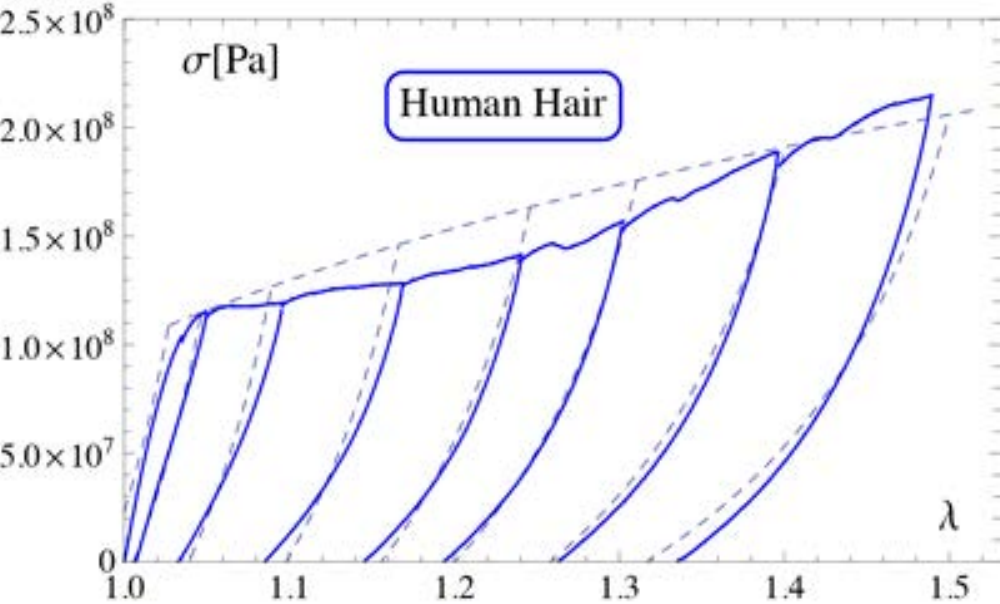}&
\includegraphics[scale=0.50]{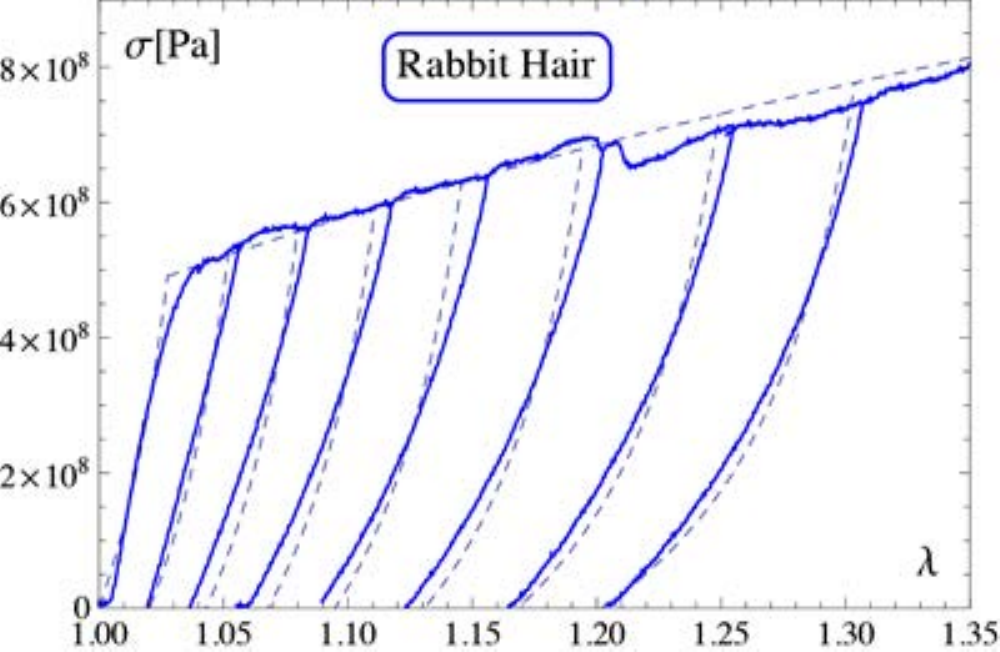} &
\includegraphics[scale=0.50]{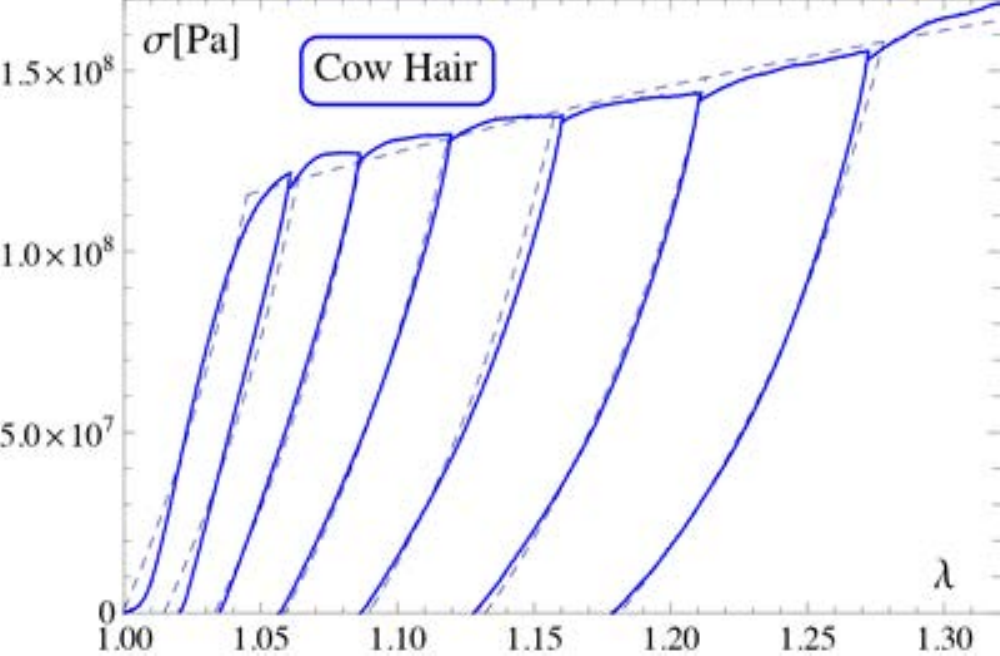}\vspace{0 cm}\\
\mbox{a)} &\mbox{b)} &\mbox{c)}\\
\end{array}$$
  \caption{
 Cyclic tests on human a), rabbit b), and cow c) hair  at room temperature and at a strain rate of 0.1 s$^{-1}$. For the theoretical model we used in a) for human hair
 $l_p = 0.4$ nm, $T = 300$ K, $q=Q/l_c=45 $ $k_B T$/nm, $\mu=  44$ MPa,  $p = -130$ MPa, $N^{fib}=10^{18}$; 
in b) for rabbit hair $l_p = 0.45$ nm; $T = 300$ K, $q=Q/l_c=180$ $k_B T$/nm, $\mu=280$ MPa,  $p = -375$ MPa, $N^{fib}=10^{18}$;  in c) for cow hair
     $l_p = 0.33$ nm; $T = 300$ K, $q=Q/l_c=45$ $k_B T$/nm, $\mu=200$ MPa,  $p = -130$ MPa, $N^{fib}=10^{18}$.
 \label{kerker} }\vspace{-0.04 cm}
\end{figure}

We first consider {\bf keratinous materials},  such as human, cow, and rabbit hairs. The tests and the possibility of reproducing the macroscopic behavior of these keratin fibers are shown in Fig.~\ref{kerker}.

\end{multicols}

\begin{multicols}{2}

 The macromolecular parameters are coherent with the values reported in the literature.  The values of the persistent length match those reported in \cite{SSH} where $l_p=0.4$ nm. The values of the rate of dissipation $q=45\, k_B T$/nm agree with the value in \cite{Bu} from which one deduces that the energy for the  mechanical unfolding of a single amino acid is about $100$ kcal/mol, which  corresponds to 
$q=59.6\, k_B T$/nm, by assuming an amino acid length of about 1 nm. A higher value $q=180 \,k_B T$/nm is adopted for the rabbit hair and this is due to the  much higher experimental unfolding stress (about 600 MPa) that we here interpret as a particularly high rate of dissipation $q$ corresponding to a higher energy barrier regulating the transition between the folded and unfolded configuration. The number $N=10^{18}$ of macromolecules per unit area corresponds to a coiled-coil cross section of 1 nm as suggested in \cite{SSH}. 

 It is important to point out that the {\it only `macroscopic'} parameters of the model are the pressure $p$ accounting for the cuticle action \cite{Stan} and the shear modulus $\mu$ of the amorphous fraction. To the knowledge of the authors no value for the confinement pressure $p$ is available, whereas different methods like torsional and traverse extension and torsion tests can be adopted to evaluate $\mu$ (the values considered here are compatible with the results reported in \cite {Hearle}).  Despite the  simplification of the model  it {\bf quantitatively well reproduces} the  {\it stiffnesses}, the  {\it residual stretches}, the {\it damage softening}, and the  sudden softening corresponding to the reconnection to the primary loading curve ({\it Return Point Memory})
of the fiber whose damage and residual stretches are regulated by the previous maximum value of the stretch.

\end{multicols}

\begin{figure}[h]\vspace{-0.4 cm}\hspace{4.5 cm}
\includegraphics[scale=0.45]{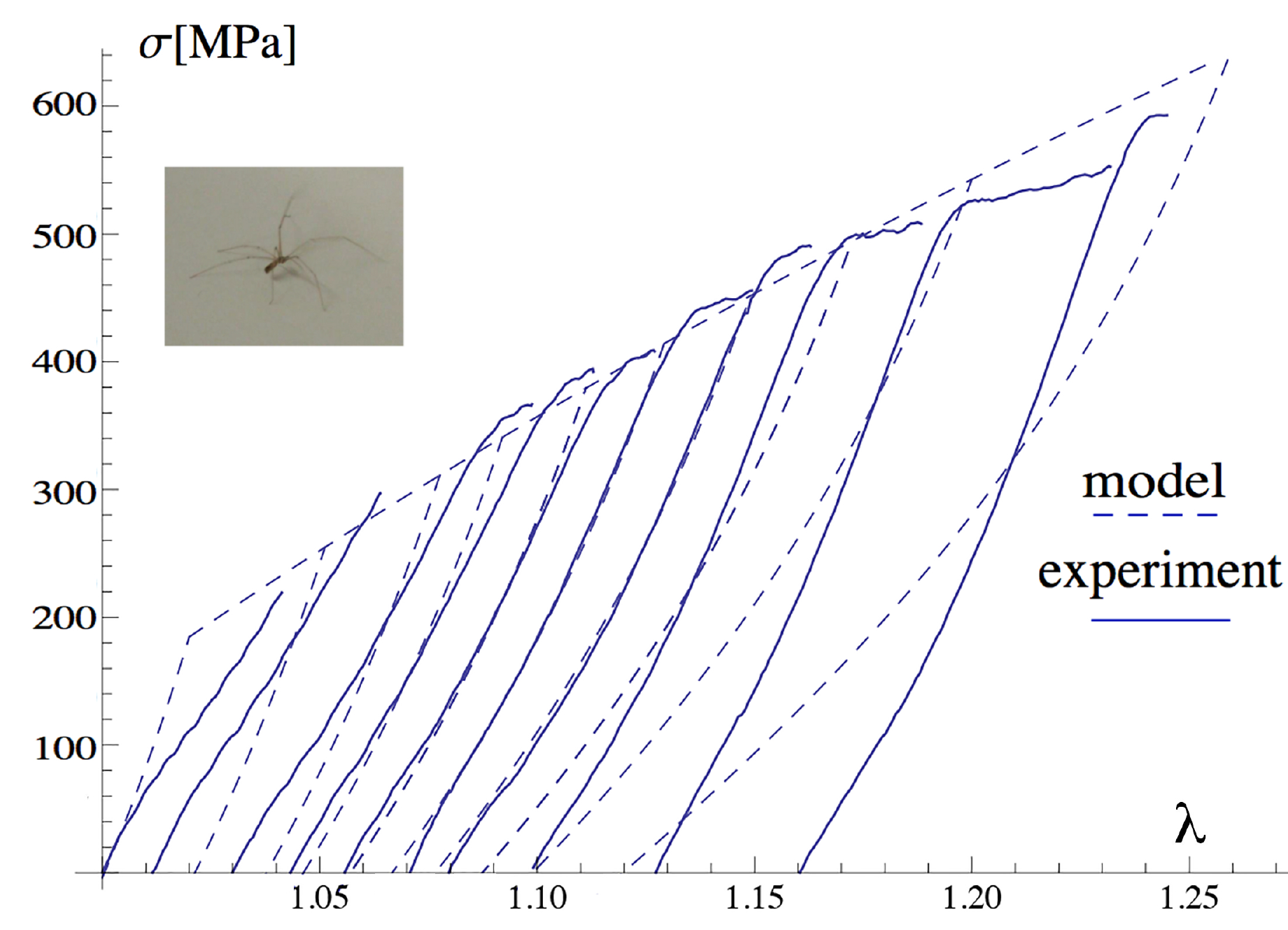}\vspace{0 cm}  \caption{ \label{espesp}
   The dragline silk sample of the spider  in the inset underwent loading/unloading cycles tested through a nanotensile machine (Agilent UTM 150) at room temperature and at a strain rate of 0.1 s$^{-1}$   (experimental behavior-continuous line, theoretical model-dashed lines). Here we assumed $l_p=0.78$  nm, $l_c=25$ nm, $N^{fib}=4.75\times  10^{18}$ m$^{-2}$, $Q=550 k_BT$, $T=300$ K, $\mu=670$ MPa, $p=-400$  MPa.
}
\end{figure}

\begin{multicols}{2}
In Fig.~\ref{espesp} we reported the cyclic behavior of a {\bf spider silk} produced by a cellar spider ({\it Pholcus phalangioides} in the inset of the figure). 
For sample preparation, the spider was left to fall down toward the ground and during its falling it produced its dragline silk, which we collected and glued on a paper frame, resulting in a fiber with 1 cm length and a diameter of about 1 $\mu$m.

The outstanding mechanical properties observed in the macroscopic material behavior is the result of a complex multiscale organization and a five layers skin-core structure \cite{Spo}. While a complete description of such behavior would ask to take care of this complex structure that extends to the microscale in a hierarchical architecture \cite{NKP}, here we focus, based on the introduced model, on the main source of dissipation, induced by the hard-soft transition \cite{Termonia} of the {\it crystalline fraction}, in form of alanine rich $\beta$-sheets, undergoing unfolding under increasing strain \cite{OS, Pap}. The skin has a confinement effect inducing a prestretch in the aligned macromolecules rich of $\beta$-sheets belonging to the core and counterbalanced by the skin pressure \cite{Pap}. Comparison of supercontracted {\it vs} native silks in \cite{Pap} determined this pressure to be about $500$ GPa. Specifically in this case the skin undergoes plastic deformation \cite{Spo} that, following again  \cite{Pap} we model as a constant  pressure $p=400$  MPa in the deformed configuration (this corresponds  to substituting the term $\frac{p}{\lambda  \sqrt{\lambda}}$ in \eqref{eeee} with $\frac{p}{\lambda}$). The matrix contribution is again modeled through a Neo-Hookean material.

Also in this case  the values of the macromolecules characteristic parameters are taken from the literature. In fact we considered the values $l_p=0.8$ nm and $l_c=25$ nm estimated in \cite{Pap,EPK} and 
\cite{OS}, the unfolding energy $Q=625 \,\, k_B T$ agrees with the values estimated in \cite{DMPS}; the number of fibers per unit area $N^{fib} = 4.75 \times 
 10^{18}$ (per square meters) reflects  the results in \cite{BKe, RV}.

While the model neglects important effects such as   the observed presence of crystal domains in the matrix \cite{Pap, BV}, the observation that the behavior at small stretches is typically regulated  by both the  amorphous fraction  and  the tertiary structure \cite{RGO, NKP}, the importance of hierarchical structure and energy exchange among different scales,  Fig.\ref{espesp} shows that it can capture at least qualitatively, but with less quantitative precision, the fundamental phenomena of damage and residual stretches. In particular we may observe the ability of estimating the dissipated energy, that for the considered spider is about 75 MJ/m$^3$ and it is in agreement with the value reported {\it e.g.} in \cite{Pap} ($130-190$ MJ/m$^3)$.

Our results show that it is possible to model with high precision the complex mechanical behavior of keratineous protein materials starting from few known molecular properties  and, even if with less quantitative precision, also the experimental behavior of spider silk. This represents in our opinion an important step forward in the  comprehension of the link between the properties of protein secondary structure and the macroscopic material properties (toughness, stiffness and large deformations prior to fracture, residual stretches), which can be of great importance for better understanding  protein materials and for designing new smart bioinspired materials.

\vspace{ 0.1 cm}
\noindent {\bf Acknowledgments}.  
NMP  is supported by the European Research Council (ERC PoC 2015 SILKENE n. 693670, by the European Commission under the Graphene Flagship (WP14 `Polymer Composites', no. 696656) and by the EU's Horizon 2020 research and innovation programme (NEUROFIBRES, no.732344).
DT and GP are supported by  funding from Italian Ministry MIUR-PRIN
voce COAN 5.50.16.01 code 2015JW9NJT.

\end{multicols}

\vspace{16 cm}
\newpage 

\newpage

\vspace{18 cm}

\begin{centering}

 {\bf SUPPORTING INFORMATIONS}
 
 \end{centering}
 
\begin{figure}[!h]\vspace{-0.05 cm }  \begin{center}\includegraphics[scale=0.45]{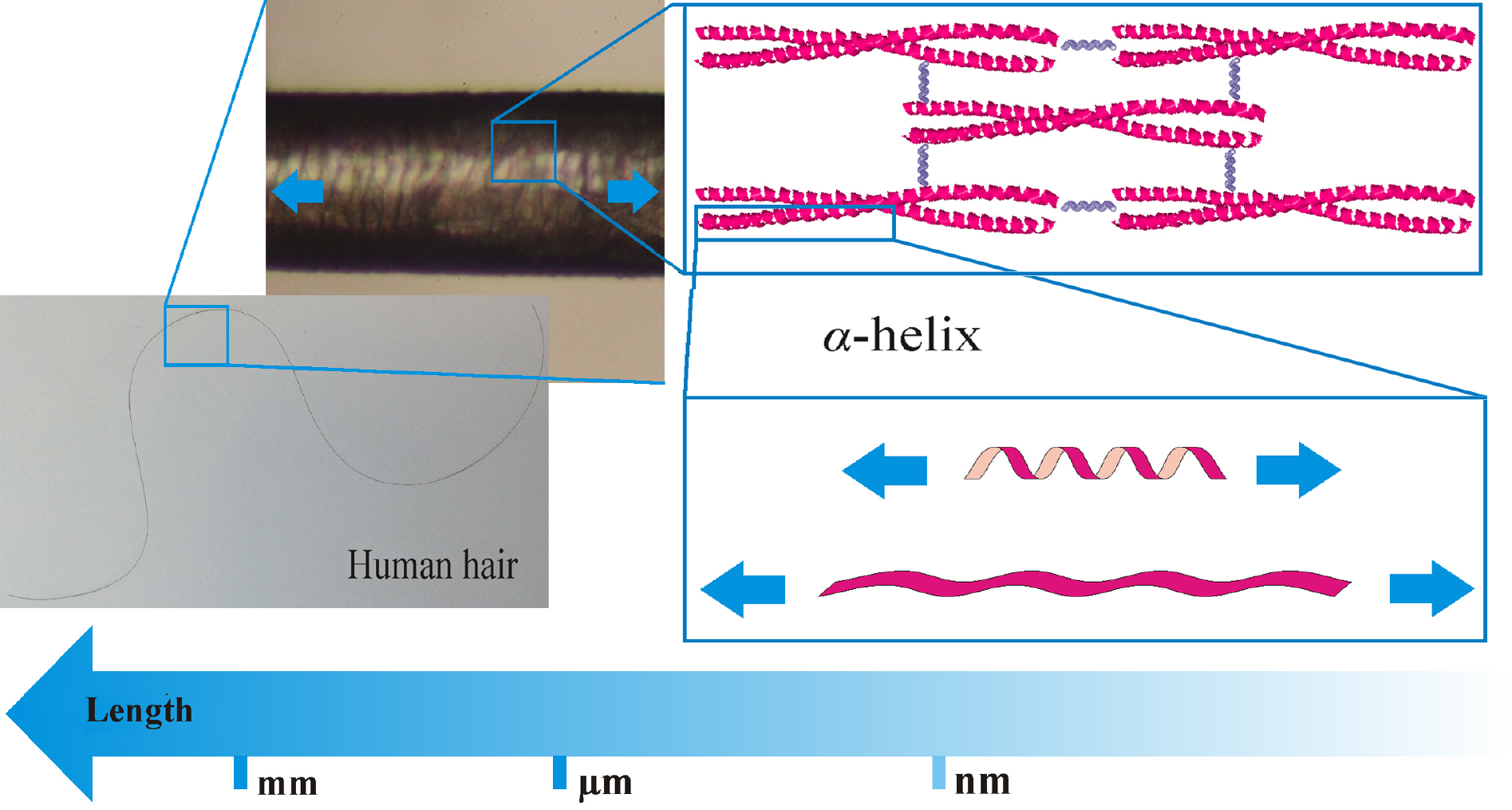} \vspace{-0.05 cm }  \caption{ \label{moms} Multiscale approach for modeling protein materials like human hair under stretching. When a hair is stretched, the $\alpha$-helix  structures undergo a coiled-coil transition to an almost straight polypeptide chain in the form of $\beta$-sheets. }\vspace{-0.2 cm}\end{center}
\end{figure}

\begin{figure}[h!]\vspace{-0.40 cm }
\hspace{3 cm}  \includegraphics[scale=0.6]{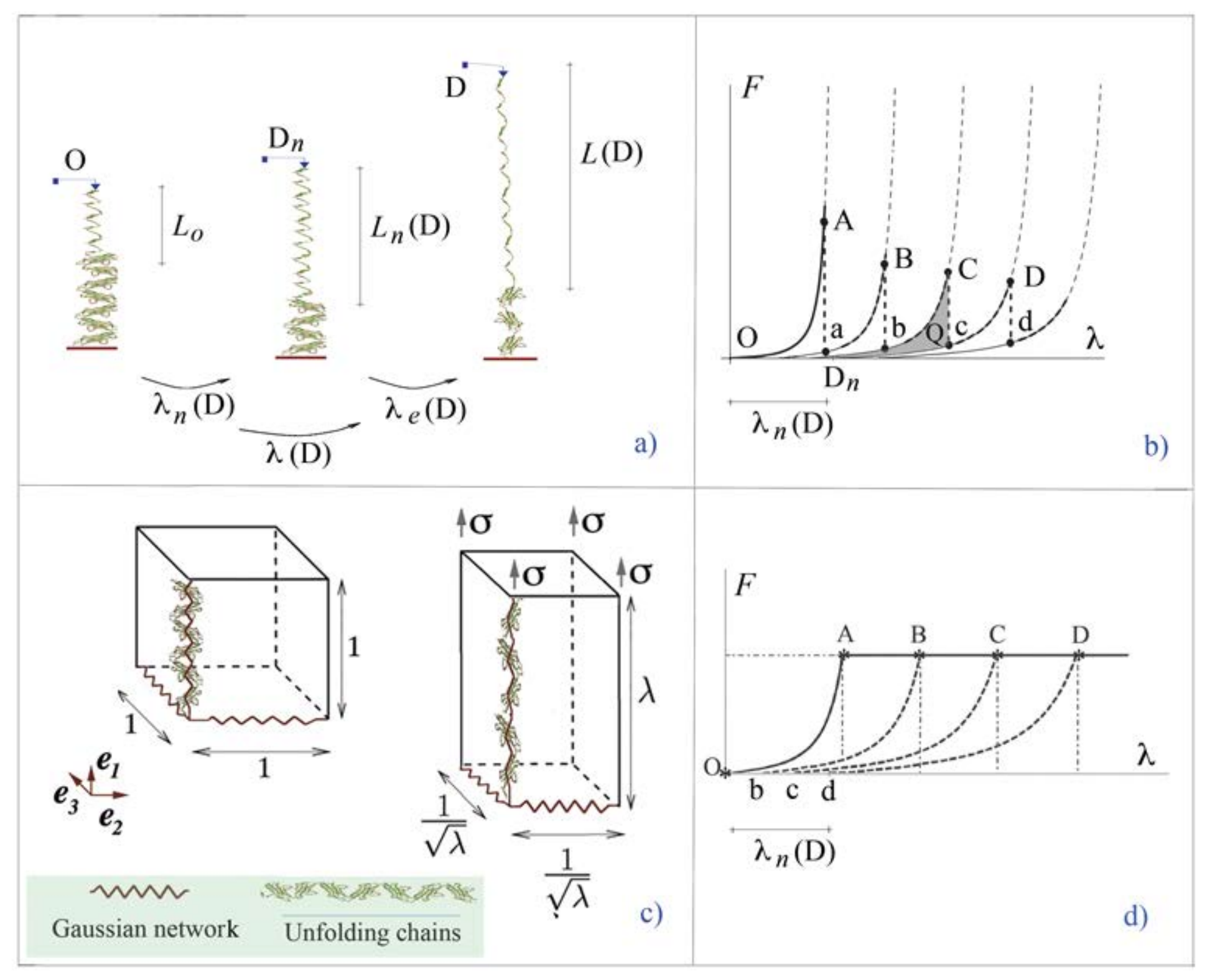} \vspace{-0.0 cm }  \caption{ \label{scheme}
a) Cartoon showing the deformation induced chain unfolding, where the total stretch $\lambda$ is given by the product of the elastic stretch $\lambda_e$ and the permanent stretch $\lambda_n$. 
 b) Force-elongation path (OAaBbCcDd) with natural length variation resulting by energy balance (path Cc). c) Micro-macro multiscale network. d) Continumm limit approximation }
\vspace{0.3 cm}
\end{figure}

\vspace{0.5 cm}
\noindent The sketch of the multiscale behavior of an air protein fiber is schematized in Fig.\ref{moms}. The \underline {main assumptions} introduced to derive our multiscale model are the following (Fig.\ref{scheme}). \vspace{0.4 cm}

\noindent {\bf a) \underline {Additive assumption}. \,\,\,} Following a classical approach in polymer elasticity \cite{FE}  (Fig. \ref{scheme}c), the energy of the  protein network is the sum of the energy of ideally non interacting unfolding chains plus a term (modeled as simple Gaussian chains) accounting for the {\it real network chains interactions}. \vspace{0.3 cm}
 
\noindent {\bf  b)  \underline {Wang and Guth hypothesis}.\,\,\,}  All macromolecules are aligned along the principal axes (Wang and Guth scheme \cite{WG}, Fig.~\ref{scheme}c)
and  undergo the macroscopic stretches ({\it affinity hypothesis} \cite{Rubinstein}).
Since protein fibers (both silk \cite{DNL} and keratin \cite{RAP}) are produced as highly oriented filaments densely packed in a less-ordered matrix, macromolecule unfolding occurs along the fiber direction, whereas the 
amorphous soft fraction is equally distributed along the  principal directions. \vspace{0.3 cm}

\noindent {\bf c)  \underline {Skin-core effect}.\,\,\,} The fiber behavior is strongly influenced by its skin-core structure (\cite{SVM}, supercontraction \cite{DLN}, confinement \cite{LSV} and prestretch of the inner chains \cite{PSK})
here considered by imposing on the fibers a transversal pressure $p$.
\vspace{0.35 cm}

\noindent \underline{Remark.} This scheme extends the model in \cite{DPSz} to consider anisotropic damage: only macromolecules along the fiber direction undergo unfolding. This is  fundamental for the experimental effect of  permanent stretches. \vspace{0.5 cm}

\noindent{\bf Macromolecule energy}
AFM experiments \cite{RGO},  schematized in Fig.\ref{scheme}a,b shows that a protein macromolecule experiences a sequence of periodic unfolding effects events as a consequence of an increasing applied displacement. This results in a contemporary  entropy and contour length increase 
({\it hidden length}: increased number of free monomers), energy `dissipation' ($Q$ in the figure) due to $H$-bonds disruption. 

Since typically  no partial unfolding occurs (either all or none of the domains unfold),
following \cite{DMPS} we model the macromolecule as a lattice of $n$ two-states (rigid-folded/entropic-unfolded domains) whose elastic energy was shown in the text to be equal to
 $\phi_e=L_c\varphi_e (\lambda_{\mbox{\tiny {\it rel}}})=
 L_c \varphi_e(L/L_c)$, where $L$ is the end-to-end length of the macromolecule, $L_c$ its contour length, $\varphi_e$ is the free energy per unit length.

Let us consider now the configurational energy of the different folded/unfolded states.
 Following \cite{DMPS} (and references therein) here we consider an Ising type transition energy 
$$\begin{array}{lll}\phi_{tr}&=& \displaystyle  -\sum_{i=1}^{n} (Q-J) (1-\chi_{i})-   J \sum_{i=1}^{n-1} (1-\chi_{i}) (1-\chi_{i+1})  \\ & 
= & \displaystyle -\sum_{i=1}^{n} Q(1-\chi_{i}) + J n_{b_f},\end{array}$$ where $n_{b_f}$ is the number of contiguous folded blocks in the folded/unfolded configuration, $Q$ is the unfolding energy of the  folded domains (considered constant) and $J$ is a penalizing `interfacial' energy term. By introducing the probability
function $p=p(L, n_u, n_{b_f})$  of a state of the macromolecule with end-to-end length $L$, $n_u$ unfolded domains, and $n_{b_f}$ continuous folded blocks, the total energy  (here $T$ is the temperature and $k_B$  the Boltzmann constant) can be expressed as
$$\begin{array}{l}\phi_{tot}= -k_BT \ln [p(L, n_u, n_{b_f})]= k_B T \Omega(n_u, n_{b_f})\exp (-\frac{\phi_{e}(L, n_u)}{k_B T})\exp (-\frac{\phi_{tr}(n_u)}{k_B T})\vspace{0.2 cm}\\ =\phi_e(L, n_u)+\phi_{tr}(n_u) -  T S(n_u, n_{b_f}),\end{array}$$
 where $\Omega(n_u, n_{b_f})$ is the number of sequences with assigned $n_u$ and $n_{b_f}$ and $S(n_u, n_{b_f})=-k_B \ln \Omega(n_u,n_{b_f})$ represents a mixing entropy component. 
 By the  di-block approximation, assuming always a single connected internal
unfolded domain inside two boundary-folded domains ($n_{b_f}=1$ as supported by the MD simulations in \cite{HS}), we neglect the mixing entropy term and the total (entropic plus unfolding)
energy is simply
\begin{equation} \phi_{tot}= \phi_e(L, n_u)+n_u Q + \mbox{const}.
\label{fitr}\end{equation}

Regarding the entropic energy of the unfolded fraction we here adopt
the elastic energy density proposed in \cite{DMPS} (this choice allows for analytic computations and keeps the same asymptotic behavior as $\l\rightarrow l_c$ of the WLC model): 
\begin{equation}\label{DPSW}
\varphi_e=\varphi_e(l,l_c)=\kappa \,\, \frac{l ^2}{l_c-l},\end{equation}
where 
$\kappa=\frac{k_BT}{4L_{p}}$ and $L_p$ is the persistence length.
Thus, using the obtained strain homogeneity result in the entropic fraction, the total elastic energy is
$\phi_e=\kappa \frac{L^2}{L_c-L} $
which allows (\ref{fitr}) to be rewritten as
\begin{equation}\label{ENERS}\phi_{\rm tot}=\kappa \frac{L^2}{L_c-L} + n_u Q.\end{equation}
 \vspace{0.2 cm}

In the spirit of the Griffith approach \cite{Gri}, following \cite{Bub}, we assume that for the given end-to-end length the configuration of the chain is attained by the minimazation of the total energy \eqref{ENERS}. The approach is schematized in Fig.\ref{scheme}b where the system undergoes an unfolding transition  (stress drop Cc in the figure) as soon as the entropic energy gain equals the (hentalpic) unfolding energy $Q$ (gray area in the figure). The efficacy of this approach for the single chain has been evidenced in \cite{DMPS}.

\vspace{0.5 cm} 

\noindent {\bf Continuum limit}
To  derive the macroscopic energy of a material consisting of a number of macromolecules, following \cite{PT} we begin by considering the  {\it continuum approximation}  of the macromolecular discrete lattice. To this end we fix the {\it total unfolded length}
$  L_c^{\mbox{\tiny 1}}=n l_c$
 and consider the limit when both  $l_c\rightarrow 0$ and $n \rightarrow \infty$ (see \cite{PT2} for a similar approach in the case of biological adhesion and   \cite{PT}, \cite{MPPT} for a rigorous thermodynamical justification of this approach). After introducing the continuous variable $x\in(0,L^1_c)$ -- such that the $i$-th link corresponds to $x\in (i\, l_c, (i+1) l_c)$, $i=1,...n$ -- and   the {\it dissipation density}  $q=Q/l_c$, the total energy for the continuum limit chain is \begin{equation}\label{Toten}
 \phi_{tot}=\phi_{tot}(L,L_c)=\kappa \frac{L^2}{L_c-L} + q L_c. \end{equation}

The equilibrium force is given by 
\begin{equation} \label{df}f=\frac{\partial  \phi_{tot}(L,L_c)}{\partial L}= \kappa\frac{2 L \, L_c -L^2}{(L_c-L)^2},\end{equation} whereas the {\it driving force} $g$ conjugated to the variation of the contour length $L_c$ ({\it i.e.} with the percentage of unravelled domains) is given by
$$g=-\frac{\partial \phi_{tot}(L,L_c)}{\partial L_c}
=\kappa\,\,\frac{L^2}{(L_c-L)^2}-q.$$ 
The Griffith approach fixes the dissipation rate $g=\hat g$, where $\hat g$ is a given material parameter here assumed null (global energy minimization).This gives for given $L_c$ the unfolding length 
\begin{equation}L_{un}\equiv L_{max}=\frac{L_c}{\sqrt{k/q}+1},
 \label{unl}\end{equation}
and a {\it constant unfolding force} (Fig.\ref{scheme}$_d$)
\begin{equation} f_{un}=\left(2
   \sqrt{k/q}+1\right)q.
   \label{unf}\end{equation}

To apply the fundamental affinity hypothesis we need to determine the evolution of chain stretches and equal them to the macroscopic ones.
We begin by observing that for a given contour length $L_c$ the {\it natural length}  (zero force)  $L_n$ of the entropic chain  can be expressed, according to a known result of Statistical Mechanics (see {\it e.g.}\!\!  \cite{Rubinstein}), as 
$$L_n=\sqrt{\bar n} b=\sqrt{\frac{L_c}{b}} \, b=\sqrt{L_c \, b} ,$$
where $b$ is the {\it length of the Kuhn segments} and $\bar n$  is the number of Kuhn segments of the chain in the present configuration. 
Thus, if we denote by $L_c^o$ the initial value of the contour length,
with $L_c^0=n_0 l_c$ and $n_0$ the initial value of the unfolded domain, we have that the initial natural end-to-end length is
$$L_o=\sqrt{L_c^o \, b}.$$
Thus
we may define the following stretch measures: 
\begin{equation}\left \{\begin{array}{ll}\lambda=\frac{L}{L_o},      & \mbox {\underline{total stretch}},\vspace{0.05 cm} \\
\lambda_e=\frac{L}{L_n},& \mbox{\underline{elastic stretch}},\vspace{0.05 cm}\\  \lambda_n=\frac{L_n}{L_o} &  \mbox{\underline{permanent stretch}},\vspace{0.05 cm}\\
 \lambda_c=\frac{L_c}{L_o} &  \mbox{\underline{limit extensibility stretch}}.\label{lam}\end{array}\right . \end{equation}
 
 We then obtain, using \eqref{df}, the force-stretch relation
 \begin{equation}\label{ffff}f\!=\!  \kappa \frac{  \frac{2  \lambda}{\bar \lambda_c(\lambda_{max})}-\left ( \frac{\lambda}{\bar \lambda_c(\lambda_{max})}\right )^2}{\left (1- \frac{ \lambda}{\bar \lambda_c(\lambda_{max})} \right )^2}\vspace{0.2 cm}\end{equation}
where $\lambda_{max}$ is the previously maximum attained strain and, by using (\ref{unf}), 
 \begin{equation}\label{lclc}\lambda_{c}=\bar \lambda_c(\lambda_{max})=\left (\sqrt{\frac{\kappa}{\bar q}}+1\right ) \lambda_{max},\end{equation}
is the stretch corresponding to the present value of the contour length.

 \vspace{0.2 cm}

The obtained continuum limit behavior is described in Fig.\ref{scheme}d. Under assigned growing end-to-end length ($\lambda_{max}\equiv \lambda$ and $\dot \lambda>0$)   the macromolecule stretches elastically until the threshold $f=f_{un}$ is attained and the macromolecule unfolds along a stress plateaux (path O-A-E). If the system is unloaded ($\lambda<\lambda_{max}$ and $\dot \lambda_{max}=0$) the system follows different paths (paths $Bb$, $Cc$, $Dd$) with permanent stretches growing with $\lambda_{max}$.

\end{document}